\documentclass[12pt,english]{article}
\usepackage[english]{babel}
\usepackage{amsmath,amssymb,amscd,color}
\usepackage{amsfonts}
\usepackage{graphicx}
\usepackage{url}

\usepackage{setspace, graphicx, amssymb, amsmath, latexsym, amsfonts, amscd, amsthm, multirow, ctable, mathdots,caption,array,diagbox,mathtools}
\usepackage{tabularx,cite,mathrsfs,subfloat,subfigure}
\usepackage{authblk}
\usepackage{color}

\usepackage[margin=2cm]{caption}

\usepackage{hyperref}

\makeatletter
\renewcommand\section{\@startsection {section}{1}{\z@}%
                                   {-5.5ex \@plus -1ex \@minus -.2ex}%nn
                                   {2.3ex \@plus.2ex}%
                                   {\normalfont\large\bfseries}}
\renewcommand\subsection{\@startsection{subsection}{2}{\z@}%
                                     {-3.25ex\@plus -1ex \@minus -.2ex}%
                                     {1.5ex \@plus .2ex}%
                                     {\normalfont\bfseries}}

\numberwithin{equation}{section}

\makeatother

\newcommand{\bea}{\begin{eqnarray}}
\newcommand{\eea}{\end{eqnarray}}
\newcommand{\be}{\begin{equation}}
\newcommand{\ee}{\end{equation}}

\newcommand{\Z}{{\mathbb Z}}
\newcommand{\R}{{\mathbb R}}
\newcommand{\C}{{\mathbb C}}

\newcommand{\M}{{\cal M}}

\newcommand{\Q}{{\mathbb Q}}

\newcommand{\cW}{{\cal W }}

\newcommand{\zb}{{\bar z}}
\newcommand{\xb}{{\bar x}}
\newcommand{\hb}{{\bar h}}

\newcommand{\nn}{\nonumber}

\def\({\left(}
\def\){\right)}

\begin{document}

		\begin{titlepage}
		
			 \begin{flushright}
\hfill{\tt CALT-TH 2020-050, IPMU20-0117}
\end{flushright}
		
			\begin{center}
				
				\hfill \\
				\hfill \\
				\vskip 0.75in
				
				{\Large 
					%\bf Moduli Spaces of 2d CFTs
					\bf On Rational Points in  CFT Moduli Spaces 
				}\\

				\vskip 0.4in
				
				{Nathan Benjamin,$^{a}$ Christoph A. Keller,${}^{b}$ Hirosi Ooguri,${}^{c,d}$ and Ida G. Zadeh${}^{e}$
				}\\
				\vskip 0.4in
				
				${}^a$
				{\it \small Princeton Center for Theoretical Science, Princeton University, 
					Princeton, NJ 08544, USA}\vskip 1mm
				${}^{b}$
				{\it \small Department of Mathematics, University of Arizona, Tucson, AZ 85721-0089, USA} \vskip 1mm
				${}^{c}$
				{\it \small Walter Burke Institute for Theoretical Physics, \\ California Institute of Technology, Pasadena, CA 91125, USA} \vskip 1mm	
				${}^{d}$
				{\it \small Kavli Institute for the Physics and Mathematics of the Universe (WPI), \\ University of Tokyo, Kashiwa, 277-8583, Japan		
				}	\vskip 1mm	
				${}^{e}$
				{\it \small International Centre for Theoretical Physics, Strada Costiera 11, 34151 Trieste, Italy} \vskip 1mm
				
				\vskip 0.4in
				
				\texttt{nathanb\_at\_princeton.edu, cakeller\_at\_math.arizona.edu, \\ ooguri\_at\_caltech.edu, zadeh\_at\_ictp.it}

			\end{center}
			
			\vskip 0.35in
			
			\begin{center} {\bf ABSTRACT } \end{center}
Motivated by the search for rational points in moduli spaces of two-dimensional conformal
field theories, 
we investigate how points with enhanced symmetry algebras are distributed there. We first study the bosonic sigma-model with $S^1$ target space in detail
and uncover hitherto unknown features. We find for instance that the vanishing of the twist gap,
though true for the $S^1$ example,  
does not automatically follow from enhanced symmetry points being dense in the moduli space.
We then explore the supersymmetric sigma-model on K3 by perturbing away from the  torus orbifold locus. Though we do not reach a definite conclusion on
the distribution of 
enhanced symmetry points in the K3 moduli space, we make several observations on how chiral currents can emerge and disappear under conformal perturbation theory.
			\vfill
		
			\noindent November 13, 2020

\end{titlepage}
\tableofcontents

\section{Introduction}

In the study of conformal field theories (CFTs) in two dimensions, particular attention has been paid to
a non-generic class of theories with a finite number of primary operators under the maximum 
chiral algebra of the theory. These special theories, called rational conformal field theories (RCFTs), 
have many interesting and simplifying properties that generic irrational CFTs do not have \cite{Friedan:1983xq}. In particular, they can in principle be solved by algebraic means. Often one is interested not just in a single CFT, but rather an entire moduli space of CFTs, which are connected by deforming with exactly marginal operators. One interesting question in this case is how RCFTs are distributed in the moduli space. For example, \cite{Gukov:2002nw} proposed conditions on rational points in the moduli space of Calabi-Yau sigma-models. Relatedly, symmetry enhancement for RCFTs is also considered relevant for understanding Mathieu Moonshine \cite{Eguchi:2010ej}.
%, Taormina:2013jza, Taormina:2013mda}.

Two classes of 2d CFTs with moduli spaces are known in the literature:
\begin{enumerate}
\item \label{class1} CFTs with a $\mathfrak{u}(1)$ global symmetry, where the exactly marginal operator is the $J\bar{J}$ operator.
\item \label{class2} CFTs with at least $\mathcal{N}=(2,2)$ supersymmetry, where the exactly marginal operators are descendants of BPS states whose dimensions are protected by supersymmetry \cite{Dixon:1987bg, Lerche:1989uy}.
\end{enumerate}
In fact, it is believed that those are the only classes with moduli spaces. The reason for this belief is that even if there are marginal fields, they will receive corrections to their weights and thus cannot serve as moduli, unless higher order contributions cancel exactly, which is not expected to happen unless the theory is either free or has enough supersymmetry \cite{Cardy:1987vr, Gaberdiel:2008fn,Behr:2013vta,Komargodski:2020ved}.
We will discuss both cases above in this paper. 

Instead of probing rational points in the moduli space directly, 
we will study a simpler question: What is the distribution of enhanced symmetry points where additional chiral fields with $\hb =0$ (or $h=0$)
emerge?  
In general,
enhanced symmetry is necessary but not sufficient for rationality. 
To characterize symmetry enhanced points, it is useful to define the twist $t$ of a
primary field as 
\be
t = 2\,\text{min}(h, \bar h)\ ,
\label{twistgap}
\ee
such that chiral fields have twist $t=0$.
If such a chiral field appears, we say that the symmetry of the theory is enhanced at that point in the moduli space. In this paper, we investigate the question whether such enhanced symmetry points are dense in the moduli space $\M$.

Let us be slightly more specific. We will assume that we know the chiral algebra $\cW_0$ that holds everywhere on $\M$. Enhanced symmetry points, where the chiral algebra becomes greater than $\cW_0$,
then form a set of submanifolds of $\M$. Some of these submanifolds may be of non-zero dimension.
For example, in the moduli space of the Calabi-Yau sigma-model, symmetries are enhanced
for toroidal orbifolds, which come with their own moduli parameters in general. One obvious situation where
enhanced symmetry points are dense is when an enhanced symmetry
submanifold is of the same dimensions as that of $\M$. However, this is not an interesting case since
we would have simply defined the enhanced symmetry as the overall chiral algebra. Thus, we assume
that there is always at least one exactly marginal perturbation from a symmetry enhanced point where  
the extra symmetry is lifted. 

In this article we investigate the density of enhanced symmetry points in two specific examples. First, in section~\ref{s:S1}, as a warm-up we consider the free boson on $S^1$ with radius $R$. Here, the entire moduli space is completely understood, and the spectrum of the theory is known explicitly in closed form: in particular the global chiral algebra is a copy of the Heisenberg algebra $\mathfrak{u}(1)$, and the theory is rational (and thus has an enhanced symmetry) if $R^2$ is rational. More generally, for toroidal CFTs $T^d$, rational CFTs can be identified in a similar way.
For $d=2$, this was worked out in \cite{Moore:1998pn} and \cite{Hosono:2002yb}, and for $d>2$, it was worked out in 
\cite{Wendland:2000ye}.

Even though this is the simplest example of the class~\ref{class1} above 
and is exactly solvable, we will see that already here certain subtleties arise.
Suppose that RCFTs are dense along a line of exactly marginal perturbation parametrized by $\lambda$,
but the point at $\lambda=0$ is not rational. 
Then, for any small $\epsilon >0$, we should be able to find a non-chiral operator $\varphi^{(0)}$ 
with weight $\hb^{(0)}>0$ at $\lambda=0$ such that the weight $\hb(\lambda)$ of
the corresponding operator $\varphi$ vanishes at some $\lambda$ where $|\lambda|<\epsilon$.
This is a striking requirement. Since $\hb(\lambda)$ has the perturbative expansion, 
\be
\hb(\lambda)= \hb^{(0)} + \hb^{(1)} \lambda + \hb^{(2)}\lambda^2 +\ldots ,
\label{hbcond}
\ee
in order to achieve  $\hb(\lambda)=0$ for some $|\lambda|<\epsilon$, 
we need $\hb^{(0)}$ to be canceled by the rest of terms in 
the series, which are seemingly suppressed by positive powers of $\lambda$. Since this should happen
for any small $\epsilon$, if there is a series $\epsilon_i$ which goes to $0$ as $i \rightarrow \infty$, 
there must be a corresponding series of  $(\varphi_i^{(0)}, \lambda_i)$ such that
$ \hb_i^{(1)} + \hb_i^{(2)}\lambda_i+\ldots $ diverges as $1/\epsilon_i$. 
In fact, we will find that there is a series of primary fields where coefficients $\hb_i^{(2)}$ diverge as $1/\epsilon_i^2$ for the free boson on $S^1$.

Another interesting observation we make is about the twist gap $t_{{\rm gap}}$ defined as the infimum of (\ref{twistgap}). It is often assumed that, if 
RCFTs are dense in the moduli space, the twist gap should vanish over the entire moduli space. 
In fact it vanishes for the free boson on $S^1$. Moveover, 
it was shown in  \cite{Benjamin:2020swg} that any CFT in  the class~\ref{class1} has a vanishing twist gap  with respect to the $\mathfrak{u}(1) \times$Virasoro algebra. 
However, we find that the vanishing of the twist gap does not automatically follow from RCFTs being dense -- there is
a logical gap in the standard argument where the twist gap $t_{{\rm gap}}$ is assumed to be a continuous function 
over $\M$.

Our second example is the non-linear $\sigma$-model on K3, where it has been conjectured that rational points are dense  \cite{Gukov:2002nw}. Here the global chiral algebra is the $c=6$, $\mathcal{N}=(4,4)$ superconformal algebra, and the exactly marginal operators are members of the $(c,c)$ and $(a,c)$ rings with appropriate conformal weights. We will start from the Kummer locus of the moduli space of K3, that is torus orbifold CFTs $T^4/\mathbb Z_2$. Extending the work of \cite{Eberle:2001jq}, we will explore the neighborhood of the orbifold locus using second order conformal perturbation theory. 

After some preliminary remarks on perturbation theory in section~\ref{s:pert}, we perturb $T^4/\Z_2$ away from the orbifold point to second order in section~\ref{s:T4orb} using a mix of numerical and analytical methods. As expected, we indeed find that all enhanced symmetry currents are lifted. More importantly, for arbitrarily small $\lambda$ we find that there is a series of primary fields such that $\hb_i^{(2)}$ diverges as $1/\epsilon_i^2$, providing some evidence that the mechanism described around (\ref{hbcond}) could indeed apply, and that enhanced symmetry points are indeed dense. This is of course a necessary, but not sufficient, condition for RCFTs to be dense in moduli space. Though we do not reach a definite conclusion whether rational points are dense in the K3 moduli space, we hope that techniques developed and observations made in this paper will 
be useful in addressing this question in future.

\section{Warm-up: $S^1$}\label{s:S1}

\subsection{Spectrum}
As a warm-up we will first consider the $c=1$ free boson theory. 
The $c=1$ free boson is of course exactly solvable, and we can write down the spectrum at all points in the moduli space. We will use a convention where $R=1$ corresponds to the self-dual point (the $SU(2)_1$ WZW model), $d^2 z$ is normalized so that the integral over the worldsheet torus gives $\tau_2$,
and the free boson $\phi$ is periodic as $\phi\equiv \phi + 2\pi$. With these conventions, the action is  given by
\be\label{Sboson}
S = \frac{R^2}{\pi} \int d^2 z \partial \phi \bar\partial \phi\ .
\ee
The partition function can then be computed as \cite{Ginsparg:1988ui}
\be
Z(\tau, \bar\tau) = \frac1{\eta(q)\eta(\bar q)} \(\sum_{n, m \in \mathbb Z} q^{\frac14\(\frac{n}R + m R\)^2} \bar q^{\frac14\(\frac{n}R - m R\)^2}\)\ .
\label{eq:pf}
\ee
Under the $\mathfrak{u}(1) \times$Virasoro algebra, the primary operators in (\ref{eq:pf}) have
\be
h^{n,m}_R = \frac14\(\frac{n}R + m R\)^2\ ,~~~~~~\bar h^{n,m}_R =  \frac14\(\frac{n}R - m R\)^2\ .
\label{eq:hhb}
\ee
Let us assume without loss of generality that $mn \geq 0$, so $h \geq \bar h$. Then we label the operator by its twist and spin:
\be
j^{n, m}_R = mn\ , ~~~~~~\frac{t^{n,m}}2=\bar h^{n,m}_R =  \frac14\(\frac{n}R - m R\)^2\ .
\label{eq:jt}
\ee
Let us now consider how the twist changes when perturbing by the exactly marginal operator $\partial \phi \bar \partial \phi$,
\be
S(\lambda) = S + \lambda \int d^2z \partial \phi \bar\partial \phi\ .
\ee
From (\ref{Sboson}) we see that this perturbation simply corresponds to a change of radius $\delta R$. Namely, if the initial radius is $R_0$ and we deform to $R := R_0 + \delta R$, then we have
\be
\pi \lambda = 2 R_0 \delta R + (\delta R)^2\ .
\label{eq:lambdaRconv}
\ee
The spin, being quantized (and $R$-independent), does not change under this deformation.  The twist changes by
\begin{align}
\bar h^{n,m}_R =  \(\frac{m R_0}2 - \frac{n}{2R_0}\)^2 &+ \(\frac{m^2 R_0}2 - \frac{n^2}{2R_0^3}\)\delta R + \nonumber\\
&+\(\frac{m^2}4 + \frac{3n^2}{4R_0^4}\) \(\delta R\)^2 + \sum_{i=3}^\infty (-1)^i \frac{(i+1)n^2}{4R_0^{i+2}} \(\delta R\)^i.
\label{eq:pt}
\end{align}
Note that the higher order terms are independent of $m$ for $i\geq3$. The sum in (\ref{eq:pt}) converges if and only if $\delta R < R_0$. 
Rewriting this in terms of $\lambda$ using (\ref{eq:lambdaRconv}), we get
\begin{align}
\bar{h}_R^{n,m} &= \(\frac{m R_0}2 - \frac{n}{2R_0}\)^2 + \frac{\pi}4\(m^2 - \frac{n^2}{R_0^4}\)\lambda + \sum_{i=2}^{\infty} \frac{(-1)^i \pi^i n^2}{4 R_0^{2+2i}} \lambda^i\nn\\
&\!\!:= \bar h_{n,m; R_0}^{(0)} + \bar h_{n,m; R_0}^{(1)}\lambda + \bar h_{n,m; R_0}^{(2)} \lambda^2 + \ldots\ .
\label{eq:ptlambda} 
\end{align}
For the free boson on $S^1$, we thus have a closed form expression for the perturbation series, which we can use as a toy model for questions about the density of rational points.

\subsection{Density of rational points}
Using this closed form expression, let us now investigate density of rational points from a perturbative point of view.
We start at a radius $R_0$, which may or may not correspond to a rational point. Given $\epsilon>0$, we want to identify a rational point within an $\epsilon$-neighborhood of $R_0$. Moreover, we also want to identify a field $\varphi$ that becomes chiral at that rational point. In view of the discussion in the introduction, we want $\varphi$ to have $\hb >0$ at the point $R_0$. 
To find a rational point near $R_0$, we simply approximate by $n,m$ such that
\be
R_0^2 + \pi \lambda =\frac nm
\label{eq:r0nm}
\ee
in such a way that $|\pi\lambda|<\epsilon$. We can then choose $\varphi$ to have  winding and momentum $(m,n)$. From (\ref{eq:hhb}) we of course immediately see that $\varphi$ becomes chiral at the rational point $R_0^2=n/m$.

How would we see this from perturbation theory though? To answer this, let us analyze the perturbative expansion (\ref{eq:ptlambda}) in more detail. Let us first point out that the terms $\hb^{(0)}$, $\hb^{(1)}$ and $\hb^{(2)}$ are special, whereas all the remaining terms satisfy
\be
\hb^{(i+1)} = -\hb^{(i)}\frac\pi{R_0^2}\ ,\quad i\ge2\ .
\ee
As long as we choose $R_0$ of order 1, that means that starting with the third order term, each higher order contribution $\hb^{(i)}\lambda^i$ is suppressed by a factor of $\lambda$. To achieve $\hb(\lambda)=0$, this means that we need
\be
\hb^{(0)} \simeq -\hb^{(1)}\lambda \qquad \textrm{or}\qquad \hb^{(0)} \simeq -\hb^{(2)}\lambda^2\ .
\ee
That is, the first or the second perturbation coefficient $\hb^{(1)}$ or $\hb^{(2)}$ has to be parametrically larger than $\hb^{(0)}$.

\begin{figure}[h!]
\captionsetup{width=1\linewidth,font = small}
\centering  
\subfigure[]{\includegraphics[width=0.49\linewidth]{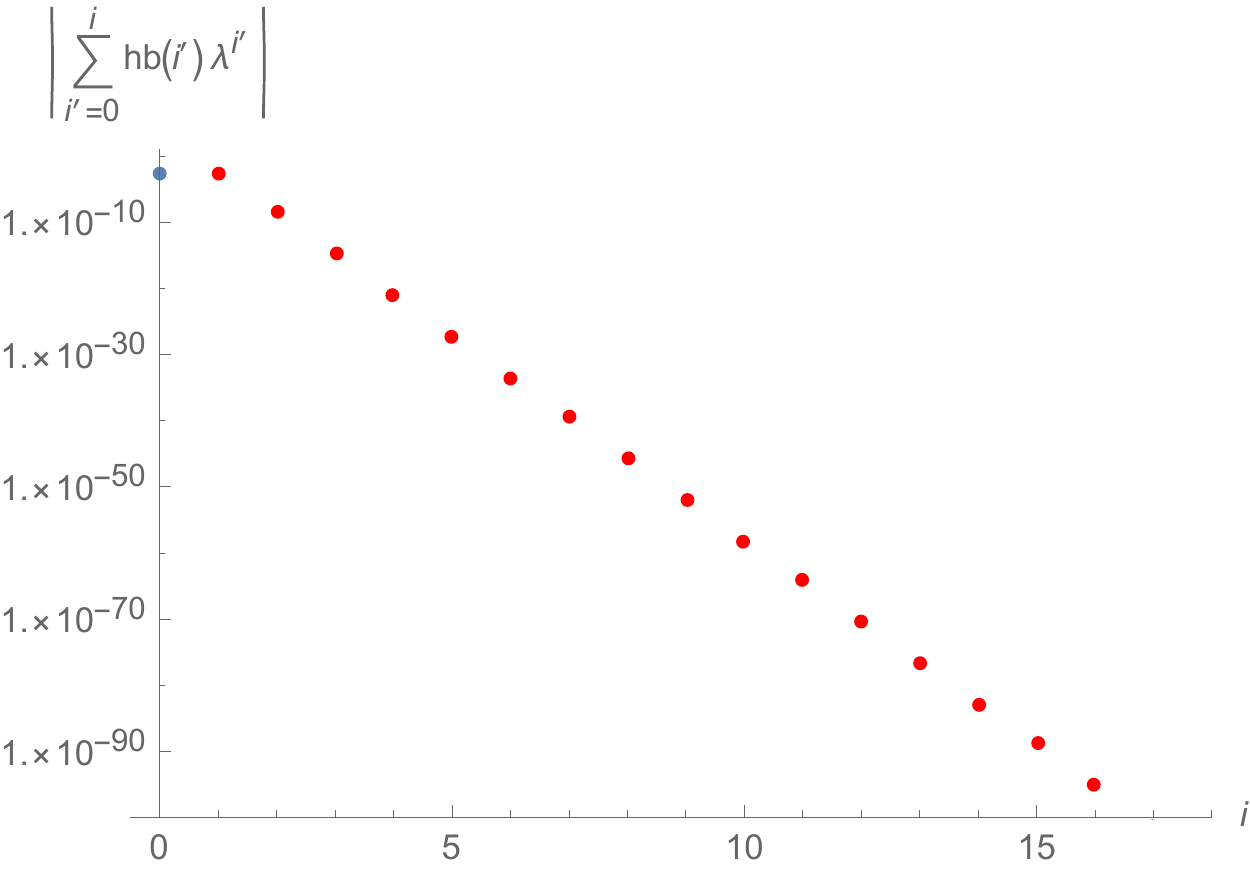}}
\subfigure[ ]{\includegraphics[width=0.49\linewidth]{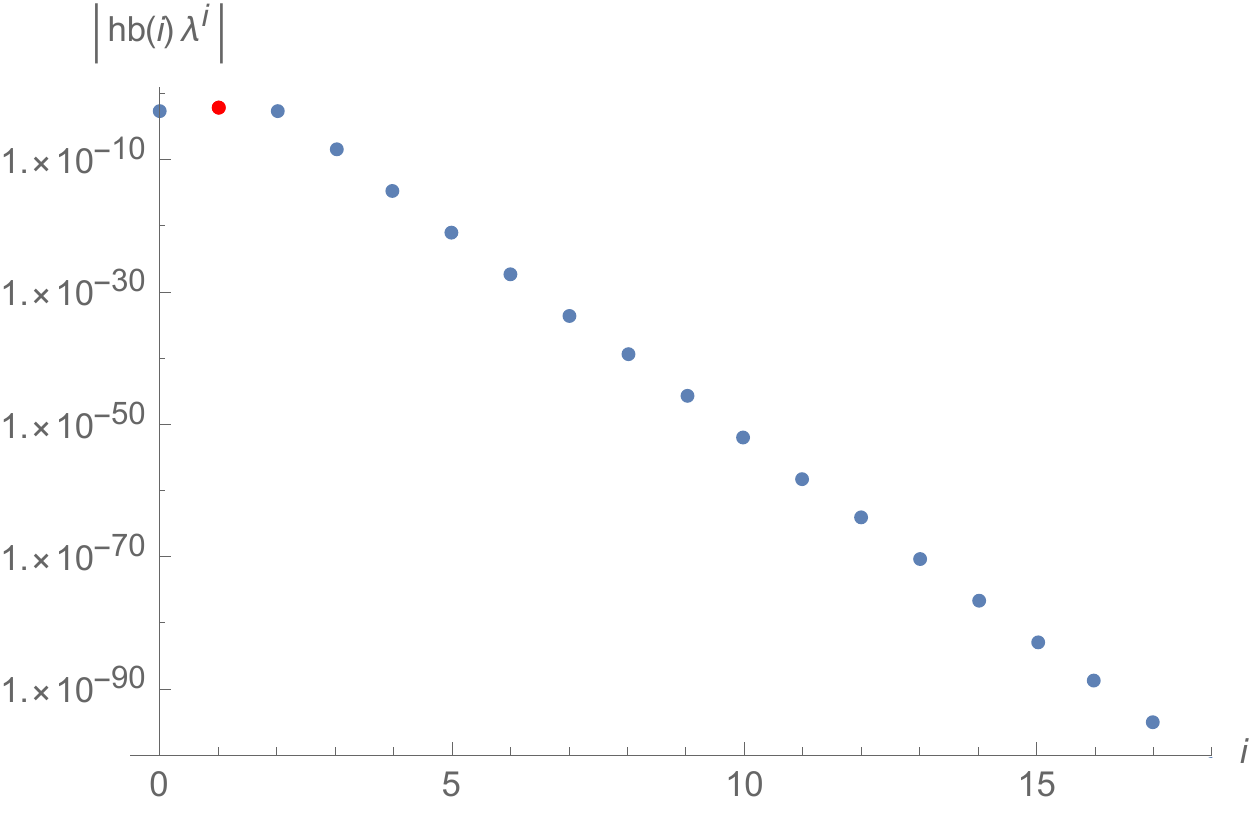}}
\subfigure[]{\includegraphics[width=0.49\linewidth]{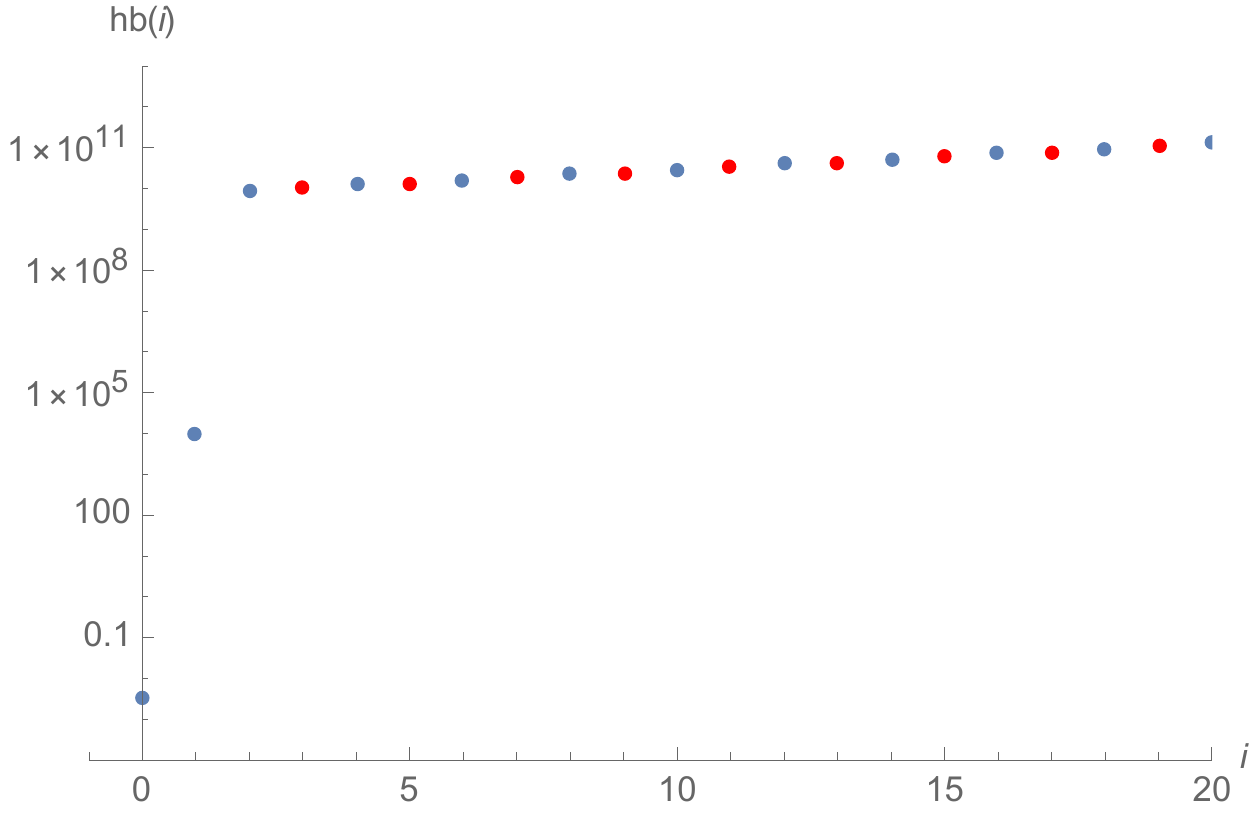}}
\caption{(a) A logarithmic plot of $\sum_{i=0}^N\bar h_{n,m; R_0}^{(i)} \lambda^i$ as a function of $N$. (b) A logarithmic plot of $\bar h_{n,m; R_0}^{(N)} \lambda^N$ as a function of $N$. (c) A logarithmic plot of $\bar h_{n,m; R_0}^{(N)}$ as a function of $N$. The zeroth, first, and second order terms in perturbation theory differ by a factor of $\lambda$ and afterwards they do not. Red and blue points indicate negative and positive numbers, respectively. 
}
\label{S1h2}
\end{figure}

To illustrate this in an example, let us take 
\be
R_0 = \sqrt{e}\ , \qquad n=271828\ , \qquad m= 100000\ .
\ee
This gives $\lambda = -5.82017\cdot 10^{-7}$ and
\be
\hb^{(0)}= 0.0031\ , \qquad \hb^{(1)} = 10566\ , \qquad \hb^{(2)} = 9.077\cdot 10^9\ .
\ee
Figure~\ref{S1h2} illustrates the growth of the coefficients and their contributions to $\hb(\lambda)$.
%Figures~\ref{fig:totalpertex1} through \ref{fig:eachpertnolambdaex1} illustrate the growth of the coefficients and their contributions to $\hb(\lambda)$.
%\begin{figure}[h!]
%\captionsetup{width=1\linewidth,font = small}
%	\centering\includegraphics[width=7cm]{TotalPertLEx1v2}
%	\caption{A logarithmic plot of $\sum_{i=0}^N\bar h_{n,m; R_0}^{(i)} \lambda^i$ as a function of $N$. Red indicates negative and blue positive. }
%	\label{fig:totalpertex1}
%\end{figure}
%
%\begin{figure}[h!]
%\captionsetup{width=1\linewidth,font = small}
%	\centering\includegraphics[width=7cm]{EachPertLEx1v2}
%	\caption{A logarithmic plot of $\bar h_{n,m; R_0}^{(N)} \lambda^N$ as a function of $N$. Red indicates negative and blue positive.}
%	\label{fig:eachpertex1}
%\end{figure}
%
%\begin{figure}[h!]
%\captionsetup{width=1\linewidth,font = small}
%	\centering\includegraphics[width=7cm]{EachPertNoLambdaLEx1v2}
%	\caption{A logarithmic plot of $\bar h_{n,m; R_0}^{(N)}$ as a function of $N$. Red indicates negative and blue positive. The zeroth, first, and second order terms in perturbation theory differ by a factor of $\lambda$ and afterwards they do not. Note the perturbation theory alternates sign which is necessary for the cancellation in Figure \ref{fig:totalpertex1}.}
%	\label{fig:eachpertnolambdaex1}
%\end{figure}

\subsection{Twist gap}
Let us now see if we can diagnose the density of rational points without needing the explicit expression for $h_{n,m}(R)$. Using our convention in eq. (\ref{eq:hhb}),
the spin is given by $j=nm$.
Next we define the \emph{twist gap} $t_{{\rm gap}}$ as
\be
t_{{\rm gap}}(R):= 2 \inf_{n,m} ( \bar h^{n,m}_R )\ .
\ee
If $R$ is rational, we clearly have $t_{{\rm gap}}(R)=0$. For irrational $R$, we can use Dirichlet's approximation theorem, which says that for any $\alpha$ and any $N$ we can find integers $m$ and $n$ with $1\leq n \leq N$ such that
\be
|n\alpha - m|  < \frac1N\ .
\ee
Picking $\alpha=R^2$, this tells us that $\hb$ can always be found smaller than $1/N$, which implies that $t_{{\rm gap}}(R)=0$ also for irrational $R$. It follows that $t_{{\rm gap}}(R)=0$ for all $R\in \R_{>0}$.

It is tempting to take the twist gap as a good diagnostics tool for determining if rational CFTs are dense in a moduli space: for instance, one could conjecture that RCFTs are dense if and only if $t_{{\rm gap}}$ vanishes everywhere on the moduli space. However, the relationship between $t_{{\rm gap}}$ and density of rational CFTs is quite subtle, and already for the simple example of the free boson, surprising things can happen.

To see this, let us generalize the twist gap $t_{{\rm gap}}(R)$ to a \emph{spin-twist gap} which we define as
\be
 \tilde t_{{\rm gap}}(R,w) := 2 ~\text{inf}\(\bar h j^w\) = \text{inf}_{n,m}\(\frac12\(\frac nR-mR\)^2n^wm^w\)\ ,
\ee
where $w$ is some real parameter.
Obviously, for any $w$, $\tilde t_{{\rm gap}}(R,w)=0$ at any rational point. Moreover, $\tilde t_{{\rm gap}}$ is also continuous at rational points: At a rational point, take $m,n$ such that $\hb_{m,n}(R)=0$. Now for these fixed $m,n$, (\ref{eq:hhb}) is clearly a continuous function of $R$, meaning for any $\epsilon>0$ we can find a $\delta$ such that for all $R'$ within $\delta$ of $R$, (\ref{eq:hhb}) at radius $R'$ is less than $\epsilon$. 
(Note that crucially $\bar{h}_{m,n}$ is not \emph{uniformly} continuous over $m$ and $n$: that is, the bigger $m$ and $n$ are, the smaller we have to choose $\delta$ to make this work.)
However, $\hb_{m,n}(R)j^w_{m,n}$ is an upper bound for $\tilde t_{{\rm gap}}(R, w)$, which implies that $\tilde t_{{\rm gap}}(R, w)$ itself is also continuous.

At irrational points however, the situation is quite different.  The Liouville-Roth constant $\mu(R^2)$ for $R^2$ is defined as the maximum value of $\mu$ such that
\be
\left|R^2 - \frac{n}{m} \right| < \frac{1}{m^{\mu}}
\label{eq:Mudefn}
\ee
has infinitely many integer solutions $n,m$. Any such $n,m$ will satisfy
\be\label{RothTwistSpin}
\tilde t_{{\rm gap}}(R,w) \leq \frac12\(\frac nR-mR\)^2n^wm^w < \frac12 R^{-2}m^{-2\mu+2+w}n^w\simeq \frac12 m^{2(1+w-\mu)}R^{2w-2}\ .
\ee
Now if $w < \mu(R^2) -1$, it follows that there will be an arbitrarily big $m$ satisfying (\ref{RothTwistSpin}), implying that $\tilde t_{{\rm gap}}(R,w)=0$. Conversely if $w \geq \mu(R^2) -1$, then all but finitely pairs $n,m$ will have $\tilde t_{{\rm gap}}(R,w)\geq \frac12 R^{2w-2}$, which means that (as long as $R^2$ is not rational) $\tilde t_{{\rm gap}}(R,w)>0$.
In summary, we have
\be
\tilde t_{{\rm gap}}(R,w) = \left\{ \begin{array}{ccc}0 &:& R^2 \in \Q \ \textrm{or}\ 
w < \mu(R^2) - 1
	\\ > 0 &:& \textrm{else}\end{array}\right. \ .
\ee
Let us work out what that means.
For $R^2$ rational we have $\mu(R^2)=1$, and by the remarks above $\tilde t_{{\rm gap}}$ vanishes and is continuous. The same also holds for certain transcendental irrational numbers called Liouville numbers \cite{Liouville}, which have $\mu(R^2) = \infty$. These Liouville numbers are dense, but have measure zero. The spin-twist vanishes and is continuous for any $w$ at rational values of $R^2$ as well as Liouville numbers. For irrational $R^2$, Dirichlet's theorem implies $\mu(R^2) \geq 2$. A very nontrivial theorem, proved by K. Roth in 1955, states that for irrational algebraic numbers, we have $\mu(R^2)=2$  \cite{MR72182}. In fact it can be shown that almost all (meaning all but a set of measure zero) real numbers have $\mu(R^2)=2$. For an introduction to aspects of this see for instance \cite{MR584443}.
 
The upshot is thus this: Even though $\tilde t_{{\rm gap}}(R,w)$ vanishes and is continuous at all rational and Liouville points, for $w\geq1$ it is nonzero at all algebraic irrational points (and indeed more generally nonzero at almost all transcendental numbers), and therefore does not vanish everywhere.

We end this subsection to emphasize again the purpose of introducing the spin-twist gap of the free boson. The spin-twist gap for $w\geq 1$ is an example of a CFT quantity that: (1) is well-defined at all points in moduli space, (2) vanishes for all RCFTs, which form a dense set of points in moduli space, and (3) does not vanish everywhere in moduli space. In principle we do not see why this phenomena could not happen for the twist gap. It therefore is logically possible for a CFT moduli space to have rational points be dense, but for the twist gap to \emph{not} vanish everywhere in moduli space (although this situation does not occur for the free boson).

\subsection{Twist gap from perturbation theory}
Although above we gave a general argument why for a free boson on $S^1$ the twist gap $t(R)$ vanishes also at irrational points, let us give an example of how this works explicitly in perturbation theory.
Suppose we start at a rational radius squared, say $R_0 = 1$, and we perturb by some irrational, small $\delta R$. The resulting theory should have an accumulation point at zero twist: for all $\epsilon >0$, there exists an operator $\mathcal{O}$ with the twist $\bar h < \epsilon$. In the limit 
\be
\epsilon \ll \delta R \ll 1\ ,
\ee
how can we see this operator in the perturbative expansion (\ref{eq:pt})?

Since this theory at the perturbed value $R$ is exactly solvable, we can explicitly see what the smallest momentum and winding modes $n, m$ are needed to get an operator with twist $\bar h < \epsilon$. To calculate this we simply use the \emph{continued fraction} representations of $R^2$:
\be
R^2 =a_0 +  \frac{1}{a_1 + \frac{1}{a_2+ \ldots}}\ .
\label{eq:cf}
\ee
Truncating (\ref{eq:cf}) gives a very good rational approximation for $R^2$. This will give an approximation of $R^2$ as a fraction, and we simply use the numerator and denominator for $n, m$. In general, the momenta and winding numbers will scale as 
\be
n, m \sim \frac1{\sqrt \epsilon}\ .
\label{eq:nme}
\ee
This is because for generic $R^2$, the Liouville-Roth constant is $2$, which means that the best rational approximation for $R^2$ (given by the continued fraction) will be 
\be
R^2 \sim \frac nm + \frac1{m^2}
\ee
from (\ref{eq:Mudefn}), which leads to an operator of twist scaling as $\frac1{m^2}$, or equivalently (\ref{eq:nme}). We pause to emphasize that it is important that we chose the rational approximation to $R^2$ ``wisely" using the continued fraction approximation. If we chose a sequence of operators with momentum and winding number $n, m$ such that the quantity $|R^2 - \frac nm|$ scaled as $\frac1m$ (instead of the optimal $\frac1{m^2}$), then this sequence would not converge to zero twist.

Similarly to (\ref{eq:nme}), the \emph{unperturbed} twist scales as
\be
%j_{n, m} \sim \frac1{\epsilon},~~ \bar h_{n,m; R_0}^{(0)} \sim \frac{\(\delta R\)^2}{\epsilon}, ~~~~~~ \frac{\bar h_{n,m; R_0}^{(0)}}{j_{n,m}} = \delta R^2
j_{n, m} \sim \frac1{\epsilon}\ ,~~ \bar h_{n,m; R_0}^{(0)} \sim \frac{\lambda^2}{\epsilon}\ , ~~~~~~ \frac{\bar h_{n,m; R_0}^{(0)}}{j_{n,m}} \sim \lambda^2\ ,
\ee
where $\lambda$ is given in (\ref{eq:lambdaRconv}). This $\epsilon$ scaling of $\bar h_{n,m; R_0}^{(0)}$ comes from (\ref{eq:nme}); the $\lambda$ scaling comes from (\ref{eq:r0nm}) and taking $\epsilon \ll \lambda \ll 1$.
Notice that the unperturbed twist is extremely large in this limit. However, each term in the perturbative expansion (\ref{eq:ptlambda}) will lower it by an additional factor of $\lambda$ until the final answer is less than $\epsilon$. 
\begin{align}
\bar h_{n,m; R_0}^{(0)} \sim \frac{\lambda^2}{\epsilon}\ ,&~~~~ \bar h_{n,m; R_0}^{(0)}+\bar h_{n,m; R_0}^{(1)}\lambda \sim -\frac{\lambda^2}{\epsilon}\ , \nn\\
\sum_{i=0}^{N}\bar h_{n,m; R_0}^{(i)} \lambda^i &\sim \frac{(-1)^N\(\pi \lambda\)^{N+1}}{\epsilon}\ , ~~~ 2\leq N < \frac{2\log \epsilon}{\log \lambda}\ ,  \nn\\
\sum_{i=0}^{N}\bar h_{n,m; R_0}^{(i)} \lambda^i &\sim \epsilon\ , ~~~ N > \frac{2\log \epsilon}{\log \lambda}\ . \end{align}

As an explicit example, we do this for the following values:
\be R_0 =1\ , ~~~\delta R = \frac{e}{10^4} \sim 2.7 \times 10^{-4}\ , ~~~\epsilon = 10^{-35}\ .\ee
These values give $\lambda \sim 1.7 \times 10^{-4}$ from (\ref{eq:lambdaRconv}). 
For the example we considered above, it turns out we need to keep up to $a_{29}$ in the continued fraction to get a precision within $10^{-35}$. This gives 
\be
n=451804615502004031\ , ~~~m=451559089162740294\ .
\ee 

\begin{figure}[h!]
\captionsetup{width=1\linewidth,font = small}
	\centering\includegraphics[width=7cm]{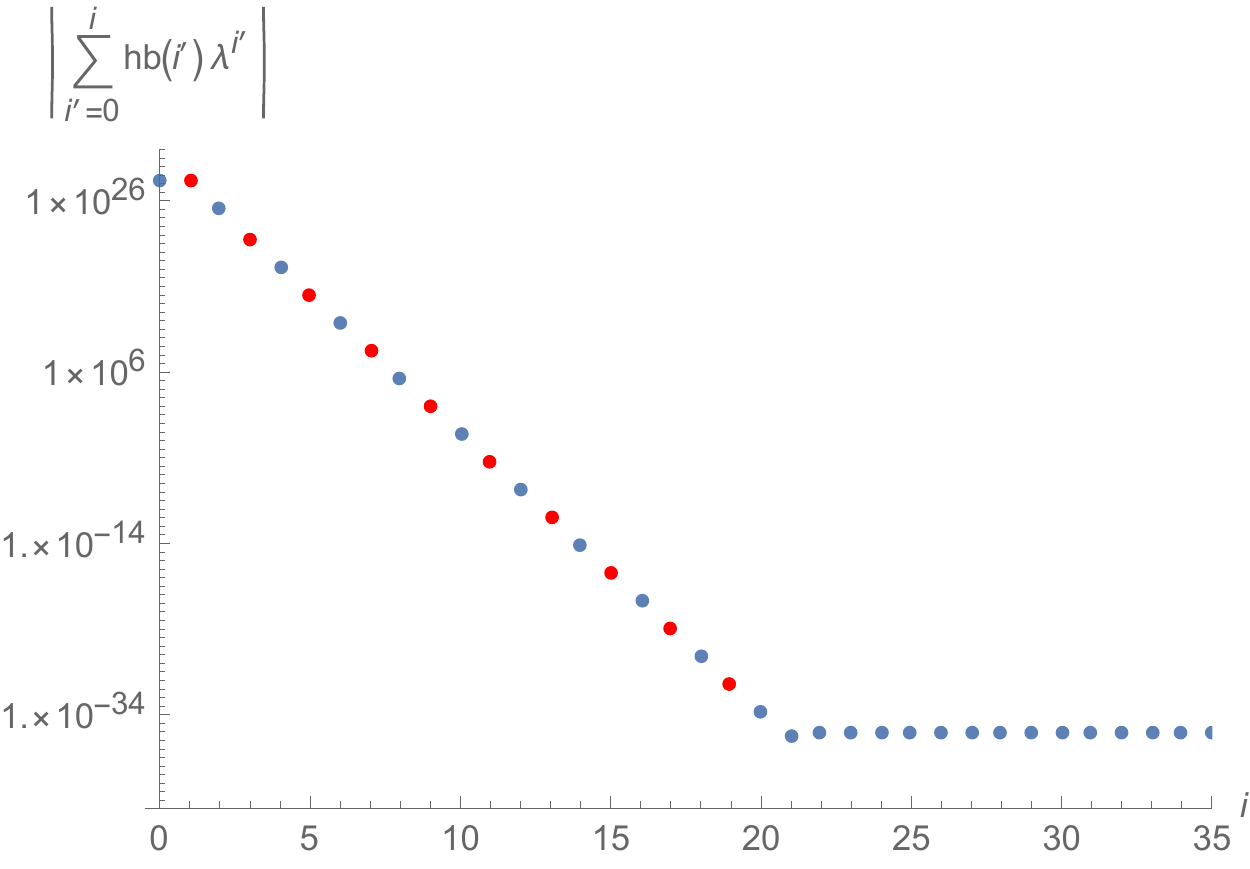}
	\caption{A logarithmic plot of $\sum_{i=0}^N\bar h_{n,m; R_0}^{(i)} \lambda^i$ as a function of $N$. Red indicates negative and blue positive. With each subsequent term in the perturbation theory, we lower the total twist by a factor of $\lambda$, until we reach twist less than $\epsilon$.}
	\label{fig:totalpert}
\end{figure}

In Figure \ref{fig:totalpert}, we plot $\sum_{i=0}^N\bar h_{n,m; R_0}^{(i)} \lambda^i$ as a function of $N$. The total twist after $N$ orders in perturbation theory decreases until it reaches the value below the $\epsilon$ we have chosen, when it plateaus. 

\section{Basics of perturbation theory}\label{s:pert}
\subsection{Perturbation theory in quantum mechanics}

It will turn out that conformal perturbation theory actually has many features that already appear in perturbation theory in ordinary quantum mechanics.
Let us therefore quickly review the salient points of second order perturbation theory in quantum mechanics.  We take the perturbed Hamiltonian to be $H = H_0 +\lambda V$. Let $|n\rangle$ be the eigenstates of $H_0$ with $H_0|n\rangle = E_n|n\rangle$. We will assume that
\be
\langle n | V |n\rangle =0 \qquad \forall n\ .
\ee
If there are no degeneracies, then perturbation theory gives
\be\label{QM2nd}
E_n(\lambda)= E_n + \lambda^2\sum_{m\neq n} \frac{|\langle n | V |m\rangle|^2}{E_n-E_m}
+ O(\lambda^3)\ .
\ee
We see that there is no first order correction. The second order correction tends to push $E_n$ away from energy levels close to it, and becomes more and more important as $E_n$ comes closer and closer to some other energy level $E_m$. This immediately shows us that the situation must be treated differently if there are degeneracies in the spectrum, since then the denominator diverges.

To deal with degeneracies, let us assume for concreteness that $E_n=E_m$. We then take the first order matrix
\be
\begin{pmatrix} 0&\langle n | V |m\rangle\\\langle n | V |m\rangle&0
\end{pmatrix}\ .
\ee
and choose as a basis for the degenerate subspace the orthonormal eigenvectors $|\pm\rangle$ of this matrix with eigenvalues $\pm \lambda\langle n | V |m\rangle$. Using this basis, we can  write the full second order expression 
\be\label{QMdegord2}
E_\pm(\lambda)= E_\pm \pm  \lambda\langle n | V |m\rangle
+  \lambda^2\sum_{k\notin\{+,-\}} \frac{|\langle \pm | V |k\rangle|^2}{E_\pm-E_k}
+ O(\lambda^3)\ .
\ee
The potentially divergent term has disappeared since the $|\pm\rangle$ are orthogonal eigenvectors of $V$, such that $\langle \pm| V|\mp\rangle=0$. We have thus eliminated any singular behavior at the cost of introducing a first order correction. We will see below that the same thing happens in conformal perturbation theory near so-called resonances.

%Let us also briefly remark on convergences of perturbation theory. For this we take the simplest example, namely a perturbed Hamiltonian of the form
%\be
%H = \begin{pmatrix} E_1&\lambda\\ \lambda & E_2 \end{pmatrix}\ .
%\ee
%In this case we of course do not need to do perturbation theory, but can write down the exact answer,
%\be\label{toymodel2nd}
%E_{1,2}(\lambda) = \frac{E_1+E_2\pm \sqrt{(E_1-E_2)^2+4\lambda^2}}2\ .
%\ee
%Physically of course we want to take $\lambda$ real. In the complex plane however we see that there is a branch cut at $\lambda=i\Delta E/2$. If we expand
%(\ref{toymodel2nd}) as a perturbation series in $\lambda$, we therefore expect its radius of convergence to be $\Delta E/2$. The radius of convergence is therefore given by the virtual crossing of energy levels. Physically nothing bad happens, \ie the levels do not diverge, but due to the crossing the perturbation expansion breaks down, and we would have to set up a new perturbation series.

\subsection{Conformal perturbation theory: shift of conformal dimension}
Let us now return to conformal perturbation theory  \cite{Cardy:1987vr,Dijkgraaf:1987jt,Kutasov:1988xb,Eberle:2001jq}. We are interested in the shift of the conformal weight of primary fields.
To this end we expand the two point function 
\be\label{2ptpert}
\langle\varphi(z_1)\varphi(z_2)\rangle_\lambda = \langle\varphi(z_1)\varphi(z_2) e^{\lambda\int d^2w \Phi(w)}\rangle
= \frac{1}{(z_1-z_2)^{2h(\lambda)}(\zb_1-\zb_2)^{2\hb(\lambda)}}
\ee
in powers of the coupling $\lambda$. Here $\Phi$ is an exactly marginal field. From the correlator (\ref{2ptpert}) we can read off the shift in the conformal dimensions $(h,\hb)$ of $\varphi$.  To be more precise, we are looking to compute the coefficients $h^{(n)}$ in the expansion
\be\label{hexpand}
h(\lambda)= \sum_{n=0}^\infty h^{(n)} \lambda^n\ ,
\ee
where $h^{(0)}$ is the dimension of the field in the unperturbed theory. Since the spin is integral and therefore constant under perturbations, the exact same expression (except for the $\hb^{(0)}$ term) must hold for $\hb(\lambda)$. We compute the $n^{\text{th}}$ term in the expansion of the left-hand side of (\ref{2ptpert}),
\be\label{ordern}
\frac{\lambda^n}{n!}\int d^2w_1 \ldots d^2 w_n \langle \varphi(z_1)\varphi(z_2)\Phi(w_1)\ldots \Phi(w_n)\rangle\ .
\ee
This integral is divergent due to the singularities that arise when a modulus $\Phi$ approaches another field. We can regularize it for instance by cutting out little discs of radius $\varepsilon$ around all fields.
To find the coefficients in (\ref{hexpand}), we write the right-hand side of (\ref{2ptpert}) as
\be\label{shiftexp}
\frac{1}{(z_1-z_2)^{2h^{(0)}}} \prod_{n=1}^\infty \exp(-2\log(z_1-z_2) h^{(n)} \lambda^n)\ .
\ee
For a given order in (\ref{hexpand}), the leading term in the expansion of the exponential in (\ref{shiftexp}) is given by
\be
-2\log(z_1-z_2)h^{(n)} \lambda^n\ .
\ee
%that is it comes with a minus sign. 
Higher terms in the expansion of the exponential in (\ref{shiftexp}) lead to higher powers of $\log(z_1-z_2)$. 
A shift in $h$ thus occurs if the integral (\ref{ordern}) produces logarithmic terms $\log|z_1-z_2|$. Because of dimensional analysis, such terms will always be accompanied by logarithmic terms in the regularization parameter $\varepsilon$, so it is often a convenient shortcut so simply identify terms of the form $\log \varepsilon$.

\subsection{Perturbation theory up to second order}
Let us work this out for perturbation theory up to second order.
At first order, eq. (\ref{ordern}) is of the form
\be
\lambda \int d^2w_1 \langle \varphi(z_1)\varphi(z_2)\Phi(w_1)\rangle
= \lambda 2\pi C_{\varphi\varphi\Phi} \log\left(\frac{|z_{12}|^2}{\varepsilon^2}\right) z_{12}^{-2h_\varphi}\zb_{12}^{-2\bar h_\varphi}\ .
\ee
We remove the $\varepsilon$ dependence by subtracting the counterterms.
Re-inserting the appropriate $z_{12}$ and comparing to (\ref{shiftexp}), we recover the well known result 
\be
h^{(1)} = -\pi C_{\varphi\varphi\Phi}\ .
\ee
If there are multiple fields $\varphi_i$ with the same conformal dimension, then we have to perform degenerate perturbation theory to take into account operator mixing: $h^{(1)}$ becomes now the matrix
\be
H^{(1)}_{ij} = -\pi C_{\varphi_i\varphi_j\Phi}
\ee
where $(h_{\varphi_i},\hb_{\varphi_i})=(h_{\varphi_j},\hb_{\varphi_j})$. We can then simply choose a new orthonormal basis for the fields $\varphi_i$ such that $H^{(i)}$ becomes diagonal.

We next work out the second order term. For the moment we move $z_1$ and $z_2$ to $0$ and $\infty$, respectively, and insert an orthonormal basis of states into (\ref{ordern}) to obtain
\be
\frac{\lambda^2}{2!}\sum_{\phi}\int d^2w_1 d^2 w_2 \langle \varphi|\Phi(w_1)|\phi\rangle\langle\phi|\Phi(w_2)|\varphi\rangle\ .
\ee
In this expansion, the states $\phi$ that have the same dimensions as $\varphi$ will lead to a product of two logarithmic terms. Since we chose $\phi$ to diagonalize the first order lifting matrix, the contribution is of the form
\be
\lambda^2 2\pi^2 C^2_{\varphi\varphi\Phi}\left(\log \varepsilon^{-2}\right)^2
= 2\lambda^2 (h^{(1)}\log \varepsilon^{-2})^2\ .
\ee
Once we reintroduce the $z_1$ and $z_2$ dependence, we immediately find that this is exactly the quadratic term of the expansion of the exponential of the first order shift in (\ref{shiftexp}). To get the linear term $h^{(2)}$ in the expansion of the exponential of the second order shift, we thus need to subtract the contribution of the external fields $\varphi$ in the internal channel from the correlation function. This gives a reduced correlation function that we will denote by $\langle \cdot \rangle'$.  

To compute the second order shift, we perform a M\"obius transformation
\be\label{mobius}
f(z):=\frac{(z-z_2)(w_1-z_1)}{(z-z_1)(w_1-z_2)}
\ee
which sends $w_2$ to the cross ratio $x:=f(w_2)$. The second order integral then reads
\be
\frac{\lambda^2}{2}\int d^2 w_1\,z_{12}^{-2h_\varphi}\zb_{12}^{-2\bar h_\varphi}\left|\frac{z_1-z_2}{(z_1-w_1) (w_1-z_2)} \right|^2
\int d^2 x \,\langle \varphi(\infty)\Phi(1)\Phi(x)\varphi(0)\rangle'\ .
\ee
As pointed out in \cite{Keller:2019yrr}, the $x$ integral depends on $w_1$ due to the regulator, namely the small disk cut out at the insertion point $x$. After performing the conformal map (\ref{mobius}), the radius of the disks is
\be
\frac{w_1-z_1}{w_1-z_2} \frac{\varepsilon e^{i\theta}}{z_2-z_1} + \ldots\ .
\ee

We are, however, only interested in the constant part of the $x$ integral, as it gives the coefficient of the log term, namely the shift in conformal dimensions. The answer in this case is simply given by --- see \cite[section 2.3]{Keller:2019yrr}:
\begin{align}\label{2ndorderint2}
\frac{\lambda^2}{2}\int d^2 w_1\,z_{12}^{-2h_\varphi}\zb_{12}^{-2\bar h_\varphi}\left|\frac{z_1-z_2}{(z_1-w_1) (w_1-z_2)} \right|^2
\int d^2 x \,\langle \varphi(\infty)\Phi(1)\Phi(x)\varphi(0)\rangle'\ \nn\\
= \pi\lambda^2 \log\left(\frac{|z_{12}|^2}{\varepsilon^2}\right) z_{12}^{-2h_\varphi}\zb_{12}^{-2\bar h_\varphi}
\int d^2 x \langle \varphi(\infty)\Phi(1)\Phi(x)\varphi(0)\rangle'\ .
\end{align}
This is then the `proper' contribution to the perturbation integral at second order. Reference \cite{Keller:2019yrr} argued that if power divergences in $\varepsilon$ appear, they will be removed by regularizing the integral and hence, will not contribute to $h^{(2)}$.

In total, we thus obtain
\be\label{hshift}
h(\lambda)= h^{(0)} -\pi C_{\varphi\varphi\Phi}\lambda - \frac{\pi}2 M \lambda^2 + O(\lambda^3)
\ee
where 
\be\label{H0}
M = \int_{\C} d^2 x\,G^{reg}(x)\ 
\ee
and $G^{reg}$ the integral of the suitably regulated four-point function in eq. (\ref{2ndorderint2}), or more precisely, its constant part \cite{Keller:2019suk}. 
Note that we have been somewhat cavalier in choosing our regularization scheme. In general, we would have to be more careful. However, in the cases we are interested, the first order term actually vanishes. It is then believed that the second order term is scheme independent \cite{Gaberdiel:2008fn,Behr:2013vta}.

Computation of the higher order terms in conformal perturbation theory in eq. (\ref{ordern}) entails evaluating multi-integrals of higher point functions and is beyond the scope of the present work. We, however, sketch the idea in appendix \ref{app_ordern} and in particular, briefly discuss its connection to perturbation theory in quantum mechanics.

\subsection{Weight shift near resonances}\label{ss:liftresonance}
We shall now discuss how conformal perturbation theory expression (\ref{hshift}) resembles quantum mechanics perturbation theory result (\ref{QM2nd}) at second order. Suppose we are interested in the shift of the weight of a field $\varphi_1$. 
To compute the integrated four-point function (\ref{H0}), clearly the important quantity to consider is the operator product expansion (\textsc{ope}) of $\varphi_1$ with the modulus $\Phi$.
Following \cite{Gaberdiel:2008fn}, we say a field $\varphi_2$ appearing in this \textsc{ope} is a \emph{resonance} if it satisfies
\be
h_1 + h_\Phi - h_2 =1\ , \qquad \hb_1 + \hb_\Phi - \hb_2 =1\ .
\ee
We note that in our case, since $h_{\Phi}=\hb_{\Phi}=1$, a resonance  occurs if $h_1=h_2$ and $\hb_1=\hb_2$. In particular, this means that resonances will always lead to operator mixing at first order, which means that we will actually need to perform degenerate perturbation theory at first order. However, instead of actual resonances, we will mostly be interested in \emph{near resonances}, that is fields which violate the resonance condition by a parametrically small amount $\epsilon$. As we will see, such near resonances give the dominant contribution to the integral, just like states with $E_n$ near $E_m$ gave the dominant contribution in (\ref{QM2nd}).

To evaluate the integral (\ref{H0}), we introduce the regularized correlation function
\be\label{Greg}
G^{reg}(x) = G(x)- G_0(x)-G_1(x)-G_\infty(x)
\ee
where
\be\label{4pfG}
G(x)=\langle\varphi_1(\infty)\;\Phi(1)\;\Phi(x)\;\varphi_1(0)\rangle\ ,
\ee
and the regulators $G_{x_i}(x),\ x_i=\{0,1,\infty\}$ are chosen such that the non-integrable singularities of $G(x)$ are cancelled.
Because $\Phi$ is exactly marginal, the regulator at $x=1$ contains a term of the form
\be
G_1(x)\sim  \frac1{|1-x|^4}\ 
\ee
coming from the vacuum, and possibly other terms. The singularity at $x=1$ will however not usually give a dominant contribution, so that we will neglect it for the moment. To obtain the regulator at $x=0$, we need to determine the \textsc{ope}
\be
\Phi(x)\varphi_1(0) = \frac{C_{\varphi_1\Phi\varphi_2}}{x^{h_1+1-h_2}\xb^{\hb_1+1-\hb_2}} \varphi_2(0) +\ldots\ .
\ee
Assuming that the $\varphi_2$ are orthonormal, we see that for any field $\varphi_2$ with $h_1-h_2=\hb_1-\hb_2 \geq 0$, we need to introduce a regulator term
\be\label{G0}
G_0(x) = \frac{C_{\varphi_1\Phi\varphi_2}^2}{x^{h_1+1-h_2}\xb^{\hb_1+1-\hb_2}}\ .
\ee
Similarly, we find
\be\label{Ginf}
G_\infty(x) = \frac{C_{\varphi_1\Phi\varphi_2}^2}{x^{-h_1+1+h_2}\xb^{-\hb_1+1+\hb_2}}\ .
\ee
Let us now assume that $\varphi_2$ is a near resonance. For $h_2 - h_1 = \hb_2-\hb_1 = \epsilon >0$, the field $\varphi_2$ then does not require a regulator term. Nonetheless, it will make a finite but large contribution to the integral around zero, given by
\be
\int d^2 x 
\frac{C_{\varphi_1\Phi\varphi_2}^2}{|x|^{2-2\epsilon}} \sim \frac{\pi C_{\varphi_1\Phi\varphi_2}^2}{h_2 - h_1}\ .
\ee
Likewise, $\varphi_2$ gives the same contribution to the integral around $x=\infty$. If $\epsilon$ is indeed small, we see that this contribution is large and will dominate over all other contributions. Physically, it is clear why this happens: once $h_2$ reaches $h_1$, we know that the integral becomes divergent --- after all, this is why we needed to introduce regulator terms in the first place!

Next, consider the case where the near resonance has $h_1 - h_2 = \hb_1-\hb_2 = \epsilon >0$. In this case, we need to regulate the integral by introducing $G_0(x)$ and $G_\infty(x)$. Note that now there is no major contribution of $G(x)-G_0(x)$ around 0, since $G_0(x)$ was chosen in exactly such a way to cancel out the leading contribution. However, at $x=0$ we now get a large contribution from the the regulator term $-G_\infty(x)$:
\be\label{phi1}
-\int d^2 x 
\frac{C_{\varphi_1\Phi\varphi_2}^2}{|x|^{2-2(h_1-h_2)}} \sim \frac{\pi C_{\varphi_1\Phi\varphi_2}^2}{h_2 - h_1}\ .
\ee
We see that as the weight of $\varphi_1$ moves from just below the near resonance $\varphi_2$ to just above it, we first get a positive contribution to the integral, which diverges as it approaches the resonance. It then flips sign and becomes a negative contribution. Note however that the actual divergence at the exact resonance $h_1=h_2$ is an artifact of considering second order perturbation theory: as pointed out above, in that case there is operator mixing between $\varphi_1$ and $\varphi_2$, which already gives a contribution at first order --- see (\ref{QMdegord2}) for the analogous phenomenon in quantum mechanics.

Note that so far we have only considered half of the picture: if the weight of $\varphi_1$ gets shifted, then so does the weight of $\varphi_2$. The computation of the shift is exactly the same, since if $\varphi_2$ is a near resonance of $\varphi_1$, then $\varphi_1$ is also a near resonance of $\varphi_2$. The result is thus exactly the same, except for a change of sign due to the exchange of $h_1$ and $h_2$ in eq. (\ref{phi1}). Thus, the second order shift for $\varphi_2$ reads
\be\label{phi2}
\sim -\frac{\pi C_{\varphi_1\Phi\varphi_2}^2}{h_2 - h_1}\ .
\ee
In summary, if there is a field $\varphi_2$ at near resonance to $\varphi_1$, then the integral for $\varphi_1$ has a dominant contribution
\be
M \sim \frac{\pi C_{\varphi_1\Phi\varphi_2}^2}{h_2 - h_1}\ ,
\ee
giving a shift:
\be
h_1^{(2)} \sim -\frac{\pi^2}2\,\frac{C_{\varphi_1\Phi\varphi_2}^2}{h_2 - h_1}\ .
\ee
Similarly, $\varphi_2$ is shifted by the same amount, but in the opposite direction,
\be
h_2^{(2)} \sim -\frac{\pi^2}2\,\frac{\pi C_{\varphi_1\Phi\varphi_2}^2}{h_1 - h_2}
\ee
Therefore, just as in quantum mechanics, the two operators $\varphi_1$ and $\varphi_2$ repel each other symmetrically.

\section{Perturbation theory for $T^4/\Z_2$}\label{s:T4orb}
\subsection{Four-point function}
Let us now discuss second order perturbation theory for $\sigma$ models with target space $T^4/\Z_2$. These orbifold CFTs appear at the Kummer locus in the moduli space of $K3$. We want to perturb such theories by a modulus $\Phi$ in a direction perpendicular to the Kummer locus. As such, the modulus is a blow up mode in the twisted sector of the $\Z_2$ orbifold. We compute the shift of a winding-momentum primary field with $(h, \hb )=(\tfrac{p_L^2}2,\tfrac{p_R^2}2)$. Since this field is in the untwisted sector and the modulus is in the twisted sector, the three-point function that determines the first order term vanishes, and the first non-trivial term appears at second order. This computation was discussed from a purely numerical point of view for scalar fields with small values of momentum $p$ in \cite{Eberle:2001jq}. Here we extend the computation beyond scalar fields and discuss their weight shift in the framework that we established above.

The four-point function $G(x)$ in (\ref{4pfG}) with $\varphi$ a winding-momentum field of weight $(h,\hb)$ and $\Phi$ a twist-2 modulus  was computed in \cite[section 2]{Eberle:2001jq} and is given by
\be\label{GV1V2}
G(x)=V_1^{h}(x)V_1^{\hb}(\xb)+ 
V_2^{h}(x)V_2^{\hb}(\xb)\ ,
\ee
where 
\bea\label{V1V2}
V_1^h(x) &=& \frac{4^{-2 h} x^{-h}}{(1-x)^2}\left(1+\sqrt{x}\right)^{4 h} \left(2 (2 h-1)-\frac{2 h (1+x)}{\sqrt{x}}\right)\ ,\\
V_2^h(x) &=& \frac{4^{-2 h} x^{-h}}{(1-x)^2}\left(1-\sqrt{x}\right)^{4 h} \left(2 (2 h-1)+\frac{2 h (1+x)}{\sqrt{x}}\right)\ ,\ 
%\\ V_0^h(x)&=& \frac{4^{-2 h} x^{-h}}{(1-x)^2}\ ,
\eea
and similarly for the anti-holomorphic counterparts. Note that we have
\be\label{Gcross}
G(\tfrac1x)= |x|^4 G(x)\ ,
\ee
just as we expect from crossing symmetry. 

Let us now discuss the singularities of $G(x)$ and their regularization. To regularize the singularity at $x=1$, following \cite{Eberle:2001jq} we always need to subtract the regularization term{\footnote{The two-point function of the twisted sector ground states is normalized to 1. The exactly marginal operator is constructed by acting with left and right-moving supercharges on the ground state and their two-point function is then normalized to 4, as in eq. (\ref{G1}).}}
\be\label{G1}
G_1(x)= \frac4{|1-x|^4}\ .
\ee
If $h+\hb\leq 1/2$, then we also need to subtract the term
\be
G_1(x)= \frac{4 \left(4h-1\right) \left(4\hb-1\right)}{16^{2h+2\hb} (1-x)^{2-4h} (1-\xb)^{2-4\hb}}\ .
\ee
Note that this regulator term breaks crossing symmetry (\ref{Gcross}). To find the regulator at $x=0$, we need to analyze the singularity at $x=0$. We rewrite $V_1^h$ and $V_2^h$ as
\bea
&&V_1^h(x) = -2\cdot 4^{-2 h} x^{-h} \left(\frac{\left(1+\sqrt{x}\right)^{4 h-2}}{\left(1-\sqrt{x}\right)^2}+\frac{h \left(1+\sqrt{x}\right)^{4 h-2}}{\sqrt{x}}\right)\ ,\\
&&V_2^h(x) = -2\cdot 4^{-2 h} x^{-h} \left(\frac{\left(1-\sqrt{x}\right)^{4 h-2}}{\left(1+\sqrt{x}\right)^2}-\frac{h \left(1-\sqrt{x}\right)^{4 h-2}}{\sqrt{x}}\right)\ ,
\eea
which can be expanded as
\be
V_1^h(x) = x^{-h}\sum_{n=-1}^\infty b_n(h) x^{\frac n2}\ ,
\qquad V_2^h(x) = x^{-h}\sum_{n=-1}^\infty (-1)^n b_n(h) x^{\frac n2}\ ,
\ee
where
\be\label{bnh}
b_n(h) = -2^{1 - 4 h}\bigg(h {4h -2 \choose n+1} + (n+1) {}_2F_1(2 - 4 h, -n;-1 - n; -1) \bigg)\ .
\ee
This then gives the expansion of the four-point function (\ref{GV1V2}) around $x=0$:
\be\label{G4pt}
G(x) = x^{-h}\xb^{-\hb}\sum_{n,m=-1}^\infty (1+(-1)^{m+n})b_n(h)b_m(\hb)x^{\frac n2}\xb^{\frac m2}\ .
\ee 
The leading singularity is of the form
\be
G(x) = \frac{8h\hb}{16^{h+\hb}}\frac1{x^{h+\frac12}\xb^{\hb+\frac12}}+\ldots\ .
\ee
To interpret this singularity, we note that terms of the form $x^{-h_1-h_2+h_p}$ come from fields of weight $h_p$ in the internal channel. We have $h_1 = h$ colliding with the modulus with $h_2=1$, which gives $h_p=\frac12$. As expected, the leading term in the expansion thus comes from the $\Z_2$ ground state twist field $\sigma$ which has dimensions $(\tfrac12,\tfrac12)$.
The next to leading term in the expansion of $G(x)$ at 0 is
\be\label{Gsubleading}
\frac{8}{16^{h+\hb}}\left(4 h^2-2 h+1\right) \left(4 \hb^2-2 \hb+1\right) \frac{1}{x^{h}\xb^{\hb}}\ .
\ee
This singularity comes from the field $\partial X_{-\frac12} \bar \partial X_{-\frac12}\sigma$ with dimensions $(1,1)$. This field has two (left and right-moving) bosonic descendants in the twisted sector of the orbifold.  Note that terms of the form $x^{-h-\frac12}\xb^{-\hb}$ and $x^{-h}\xb^{-\hb-\frac12}$ are absent in the expansion. This is because the angular integral in eq. (\ref{H0}) vanishes for such terms due to rotation symemtry. More generally, we note that all fields in the \textsc{ope} of $\varphi_1$ and $\Phi$ are in the twisted sector. Since there are no momentum modes in the twisted sector, this implies that resonances only occur at integer or half-integer values of $h$ and $\hb$.

\subsection{Weight shift of scalar fields}
Let us first discuss how the weights of scalar fields are shifted. We take $h=\hb=\frac{p^2}2$.
At $x=0$, the leading singularity is given by
\be\label{Leading0}
G(x) = 2 \frac{4^{-2p^2} p^4}{|x|^{p^2+1}} +\ldots\ .
\ee
We see that if $p^2<1$, this singularity is integrable and thus does not need to be regularized. By crossing symmetry (\ref{Gcross}), the leading singularity at $x=\infty$ is given 
by 
\be\label{LeadingInf}
G(x) = 2 \frac{4^{-2p^2} p^4}{|x|^{3-p^2}} +\ldots\ ,
\ee
which is again integrable. The singularity at $x=1$ is regularized by (\ref{G1}).

Let us first consider $p^2 = 1-\epsilon$ and examine the behavior of $G(x)$ in this limit. As discussed in section \ref{ss:liftresonance}, in this case the singularities at $x=0$ and $\infty$ do not need to be regularized --- see below eq. (\ref{Ginf}). The contribution of the leading term (\ref{Leading0}) near $x=0$ is given by
\be\label{divcont}
\sim \frac{2\pi}{16 (1-p^2)}\ .
\ee
By crossing symmetry, there will be the same contribution coming from the leading term at $x=\infty$.
That is, the shift diverges to $+\infty$ as $p^2 \to 1^-$. The divergence we encounter comes from the resonance $\sigma$, that is the twist sector ground state with dimensions $(\frac12,\frac12)$.

We next consider the case $p^2= 1+ \epsilon$. The singularities at $x=0$ and $x=\infty$ are now no longer integrable and must therefore be regularized --- see the discussion above eq. (\ref{phi1}). We therefore introduce the regulators
\be
G_0(x)= 2\frac{4^{-2p^2} p^4}{|x|^{p^2+1}}\ , \qquad 
G_\infty(x) = 2\frac{4^{-2p^2} p^4}{|x|^{3-p^2}}\ .
\ee
We now compute the contribution of $G^{reg}(x)$ (\ref{Greg}) around $x=0$. By construction, the integrand only has an integrable singularity, since (\ref{Leading0}) gets precisely canceled by $G_0(x)$. There is therefore no contribution of the form (\ref{divcont}). However, there is an important contribution coming from the regulator $G_\infty(x)$ --- see eq. (\ref{phi1}). We namely have
\be
\sim  \frac{2\pi}{16(1-p^2)}\ .
\ee
From crossing symmetry, we expect the same contribution at $x=\infty$. This explains why the shift diverges to $-\infty$ even as $p^2 \to 1^+$.
As we increase $p^2$, we expect the same to happen at all integer values. For instance, (\ref{Gsubleading}) shows that there will be a resonance at $h=\hb =1$, coming from the field $\partial X_{-\frac12} \bar\partial X_{-\frac12}\sigma$. 

\begin{figure}[htbp]
	\centering
\captionsetup{width=1\linewidth,font = small}
	\includegraphics[width=0.5\textwidth]{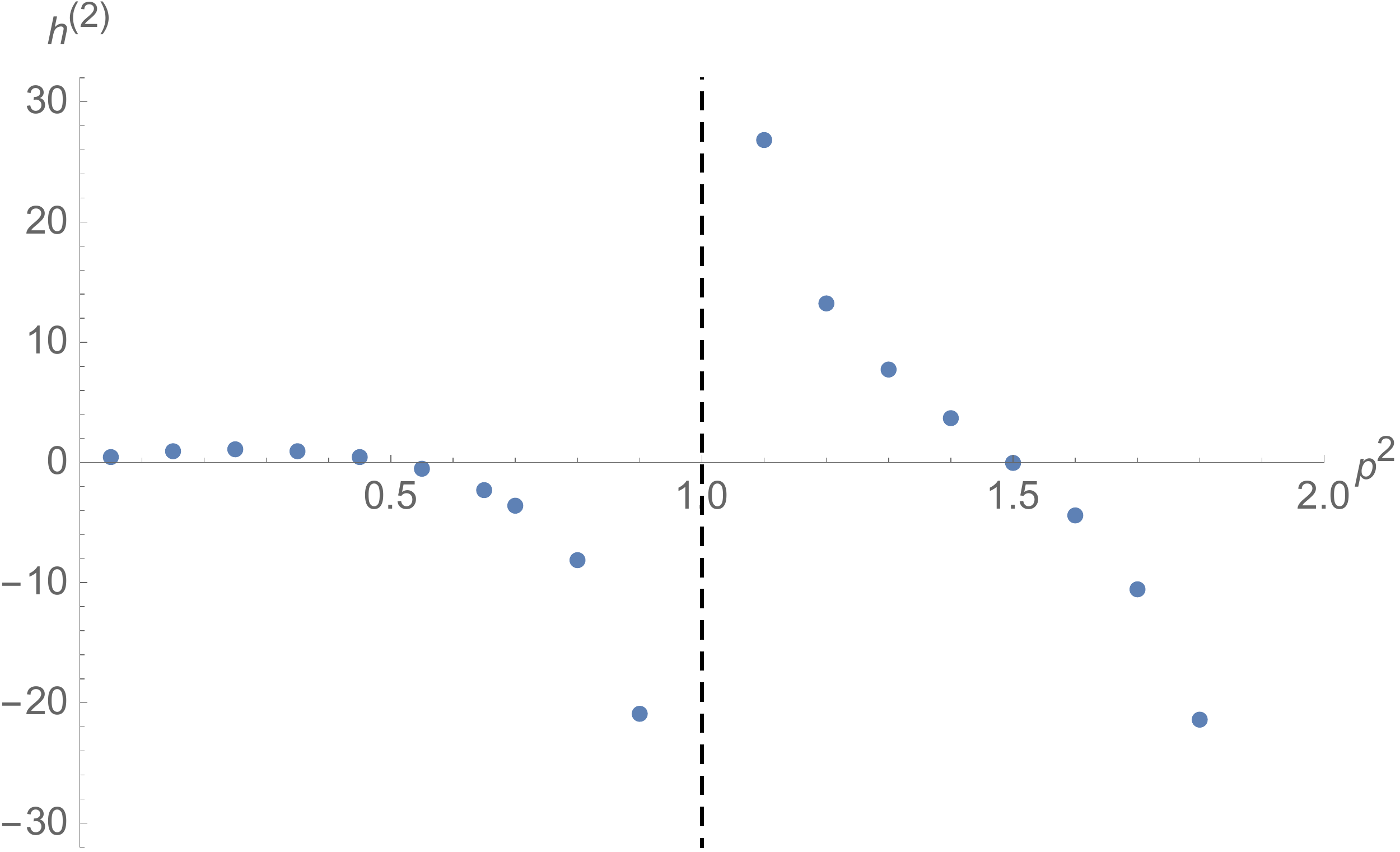}
	\caption{Second order shift $h^{(2)}$ as function of momentum $p^2$ for scalar fields. The first resonance occurs at $h=\hb=\tfrac12$.}
	\label{f:scalarlifting} 
\end{figure}

The above analysis is indeed borne out when we compute the second order shift $M$ (\ref{H0}) by evaluating the integral numerically. To do this we find it useful to cut out small disks around 0, 1  and $\infty$. We then perform the integral numerically outside of those disks, avoiding numerical instabilities coming from the cancellations between $G(x)$ and the regulators. To obtain the contributions from those discs, we perform a series expansion of the regulated correlation function and then integrate the first few terms. We have plotted the outcome in figure~\ref{f:scalarlifting}. We note that our results agree with \cite{Eberle:2001jq} for the values of $p^2$ that were computed there.

\subsection{Weight shift of spin 1 fields}
\begin{figure}[htbp]
\captionsetup{width=1\linewidth,font = small}
	\centering
	\includegraphics[width=.5\textwidth]{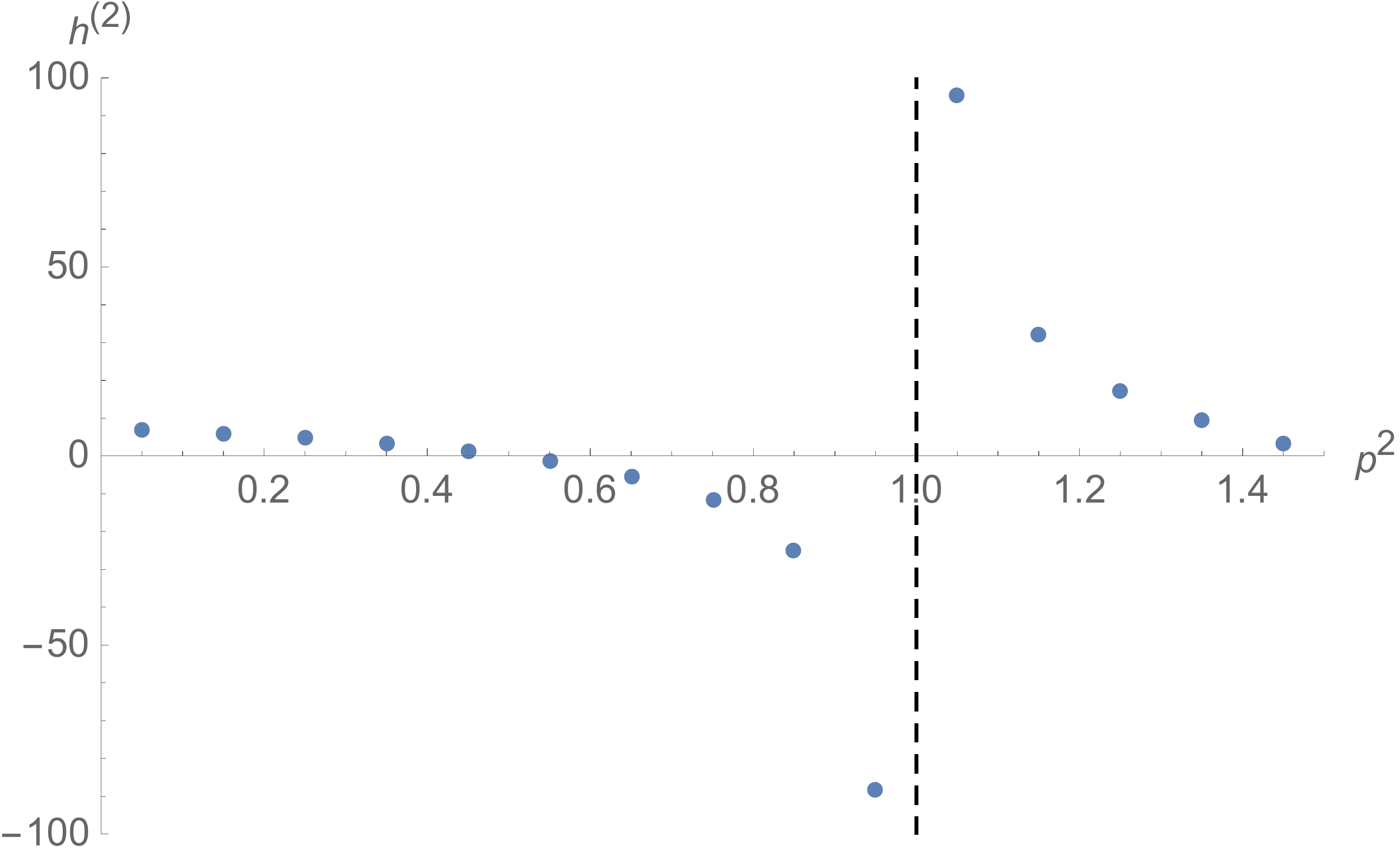}
	\caption{Second order shift $h^{(2)}$ as function of $p^2$ for spin 1 fields with $(h,\hb)=(\frac{p^2}2,\tfrac{p^2}2+1)$ or $(\frac{p^2}2+1,\tfrac{p^2}2)$. The first singularity occurs at $p^2=1$. }
	\label{f:spinlifting} 
\end{figure}

We next consider fields with spin 1. Note that the leading term in the expansion of the four-point function (\ref{G4pt}) only contributes to the integral if $h=\hb$, since otherwise the angular integral vanishes --- see the comment below eq (\ref{Gsubleading}). As such, the leading term in eq. (\ref{G4pt}) does not contribute to the weight shifting of non-scalar fields.
The first term in the expansion contributing to the shift of spin 1 fields is
\be\label{spin1h2}
\frac{8h \hb}{16^{h+\hb}}  (7-10 h+8 h^2) \frac1{x^{h+\frac12}\xb^{\hb-\frac12}}
\ee
as well as a similar term with $x \leftrightarrow \xb$. Eq. (\ref{spin1h2}) corresponds to the contribution from an internal field of the form $\bar\partial X_{-1/2}\bar\partial X_{-1/2}\sigma$ or $\bar\psi_{-1}\sigma$. This means that for a spin 1 field, the first resonance will occur at $h=\tfrac12, \hb=\tfrac32$. This is again borne out when we compute the second order shift $M$ (\ref{H0}) numerically, as can be seen in figure~\ref{f:spinlifting}.

\subsection{Lifting chiral fields}
Finally let us discuss the lifting of chiral fields with dimensions $\hb=0, h \in \Z_{\geq 0}$. 
What we want to check here is that indeed none of the extra currents at the $\mathbb Z_2$ orbifold point survive perturbation. This is closely related to the results found in  \cite{Gaberdiel:2015uca,Keller:2019suk,Keller:2019yrr}. From eqs. (\ref{GV1V2}) and (\ref{V1V2}) we have
\be
G(x)= \frac{-2}{(1-\xb)^2}(V^h_1(x)+V^h_2(x)) 
=\frac{-2}{(1-\xb)^2 x^h} \sum_{n=0}^\infty 2b_{2n}(h)x^n\ .
\ee
Since $G(x)$ has such a simple $\xb$ dependence, we can evaluate the integral analytically. Rather than subtracting divergences, we will instead use the original regularization scheme of cutting out $\varepsilon$-discs around $0,1,\infty$. We use Stokes' theorem to perform the $\xb$ integral.
Namely we have 
\be
\int_{\partial U} F dz + G d\zb = \int_U \left(\partial_z G- \partial_\zb F\right) dz d\zb
\ee
where the complex integration measure is
\be
dx\wedge dy = \frac{i}{2} dz \wedge d\bar z\ .
\ee
The integral of $G(x)$ over the complex plane (\ref{H0}) then gives contour integral
\be
M_h = -i \oint_{0,1,\infty} dx \frac1{(1-\xb)} (V^h_1(x)+V^h_2(x))\ .
\ee
Note that the contours around $0$ and $1$ are counterclockwise now, which has led to an additional minus sign.
As expected, we do not get a contribution from $1$, the location of the other modulus. More precisely, we do get a term $\varepsilon^{-2}$, which is regularized away, but no constant term.
The integral around $x=0$ gets a contribution from the $n=h-1$ term only, and evaluates to
\be
4\pi b_{2h-2}(h)\ .
\ee
The integral around $x=\infty$ gives the same contribution, so that in total we get
\be\label{Mchiral}
M_h = 8\pi b_{2h-2}(h) = -\frac{4\sqrt{\pi} \Gamma(2h+\frac12)}{\Gamma(2h)}.
\ee
For the first few values of $h$, the values for the second order lifting $h^{(2)}=-\frac{\pi}2 M_h$ is given in table \ref{chiralh2}.
\begin {table}[h!]
\captionsetup{width=1\linewidth,font = small}
 \begin{tabular}{|c||c|c|c|c|c|c|c|c|c|} 
 \hline
 $h$ &0&1&2&3&4&5&6&7&8 \\% [0.5ex] 
 \hline 
$h^{(2)}$&0&$\frac{3 \pi ^2}{2}$&$\frac{35 \pi ^2}{16}$&$\frac{693 \pi ^2}{256}$&$\frac{6435 \pi ^2}{2048}$&$\frac{230945 \pi ^2}{65536}$&$\frac{2028117 \pi ^2}{524288}$&$\frac{35102025 \pi ^2}{8388608}$&$\frac{300540195 \pi ^2}{67108864}$\\[0.5ex] 
 \hline
\end{tabular}
\caption{Second order lifting $h^{(2)}$ for chiral fields with dimension $(h,0)$.}\label{chiralh2}
\end{table}

\begin{figure}[htbp]
\captionsetup{width=1\linewidth,font = small}
	\centering
	\includegraphics[width=.5\textwidth]{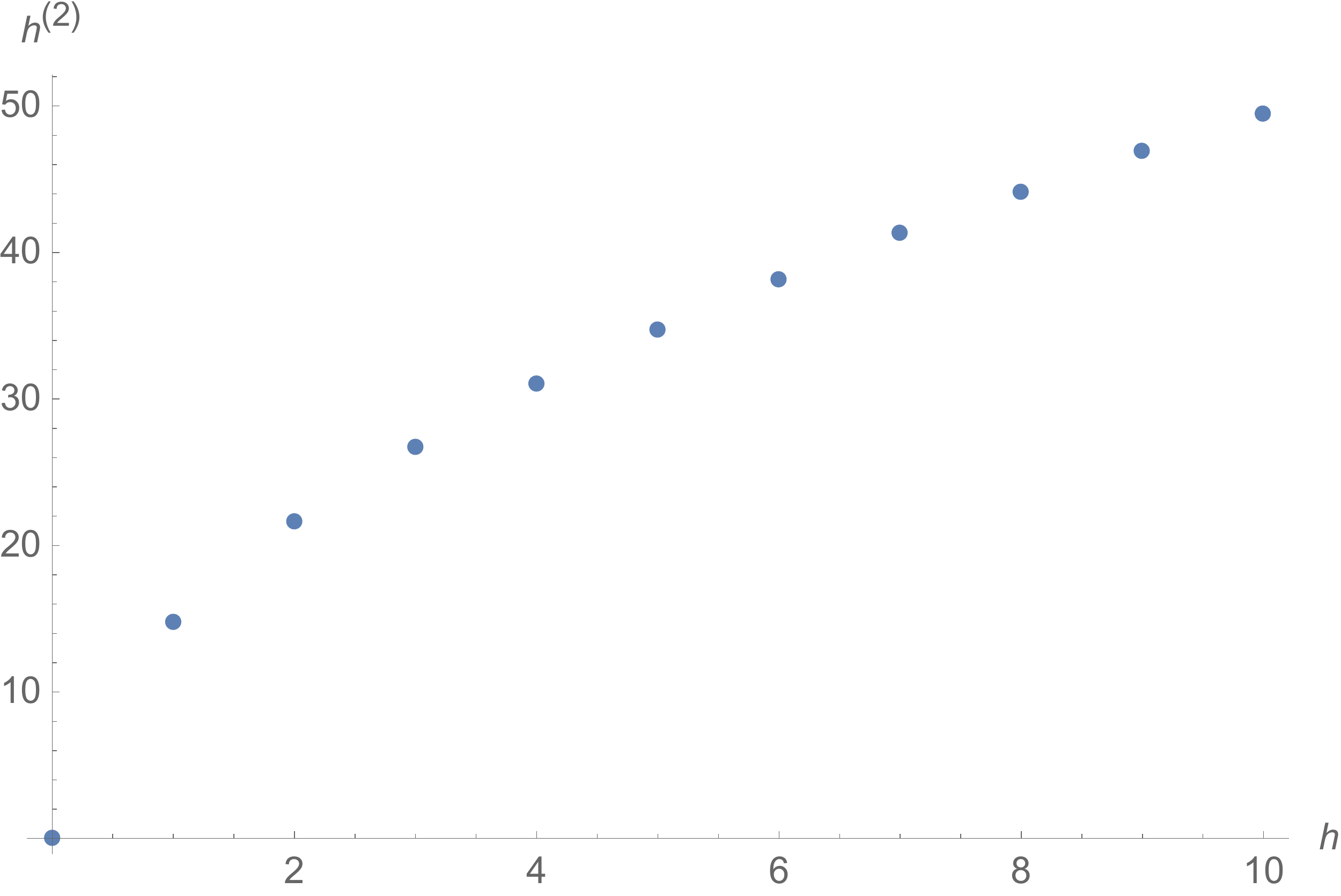}
	\caption{Second order lifting $h^{(2)}=-\frac{\pi}2M$ for chiral fields as function of $h$ --- see eq. (\ref{Mchiral}).}
	\label{f:chirallifting} 
\end{figure}

For $h>0$ the lifting is always strictly positive, which then shows that all chiral fields get lifted. We have plotted lifting of chiral fields in figure~\ref{f:chirallifting}. We see that they grow in absolute value with increasing $h$. In fact, we have
\be
h^{(2)} = (2\pi)^{3/2}\sqrt{h} + O(h^{-1/2}).
\ee
This agrees with our general hypothesis that perturbation theory converges worse with increasing dimension.

\subsection{General resonances}
We now discuss shifting of general winding-momentum primaries of the $T^4/\mathbb Z_2$ orbifold theory and examine whether we can identify candidates that at least in principle could become new chiral fields. This then signals the possibility of the existence of a rational point infinitesimally close to the Kummer surface and perpendicular to it in the K3 moduli space. We start out with a non-rational torus. This means that we can find an arbitrarily small $\epsilon>0$ such that there exists a field $\varphi$ with dimensions
\be
(h,\hb)=(j+\tfrac12-\epsilon,\tfrac12-\epsilon)\ .
\ee 
%for small $\epsilon$ and large spin $j$.
From eq. (\ref{G4pt}) we see that $\varphi$ is close to the resonance for $n=2j-1$ and $m =-1$, and contributes to the four-point function as
\be
2b_{2j-1}(h)b_{-1}(\hb) \frac1{x^{1+\epsilon}\xb^{1+\epsilon }}\ .
\ee
For $h \simeq j+\tfrac12$ and $\hb\simeq\tfrac12$ we have
\be\label{bs}
b_{-1}(\tfrac12)=-\frac14\ ,\qquad b_{2h-2}(h) = -\frac{\Gamma(2h+\frac12)}{2\sqrt{\pi}\,\Gamma(2h)}= -\frac{\sqrt{h}}{\sqrt{2 \pi }} +\ldots\ .
\ee
In total, we thus get
\be
\hb^{(2)}\lambda^2 \sim -\frac{\lambda^2}{\epsilon}\,\frac{1}{2}\,\sqrt{\frac{\pi }{2}}\,\sqrt{j+\tfrac12}\ .
\ee
This means that no matter how small we choose $\lambda$, we can always find a field $\varphi$ with $\hb\simeq\tfrac12$ and $j$ large such that $h^{(2)}\lambda^2$ is negative and of order 1, meaning
\be\label{h2candidate}
\hb^{(2)}\lambda^2 \simeq -\hb^{(0)}\ .
\ee 
The second order correction thus has a chance of shifting $\hb$ close to zero, which make this field a natural candidate for a new chiral field. 
Eq. (\ref{h2candidate}) of course does not guarantee that: since the second order term is of order 1, it tells us that we need to go to higher order perturbation theory. Still, it is encouraging to find that such candidate fields always exist.

Note that even if we are at a rational point, where all fields have weight at least $\epsilon$ removed from half-integers, we can still find a field for which (\ref{h2candidate}) holds simply by choosing $j$ large. Also note that the same argument works for $\epsilon<0$. In that case the weight of $\varphi$ increases. However, by the remarks at the end of section~\ref{ss:liftresonance}, the weight of the resonance then shifts downwards by the same amount. This simply means that in this case the candidate field is not $\varphi$, but rather the resonance in the twisted sector.

\section*{Acknowledgements}
We thank C.~Bachas, D.~Gepner, K.~Narain, C.~Vafa, and K.~Wendland for interesting discussions.  
The work of N.B. is supported in part by the Simons Foundation Grant No. 488653. 
The work of C.A.K. is supported in part by the Simons Foundation Grant No.~629215.
The work of H.O. is supported in part by
U.S.\ Department of Energy grant DE-SC0011632, by
the World Premier International Research Center Initiative,
MEXT, Japan, by JSPS Grant-in-Aid for Scientific Research 17K05407  and 20K03965,
and by JSPS Grant-in-Aid for Scientific Research on Innovative Areas
15H05895.
N.B. and H.O. thank the Aspen Center for Theoretical Physics, which is supported by
the National Science Foundation grant PHY-1607611,  where part of this work was done. 

\appendix
\section{`Connected' perturbation theory} \label{app_ordern}
In this appendix we briefly discuss the contribution of higher order terms in conformal perturbation theory which were introduced in eq. (\ref{ordern}). We discuss some ideas of how reduced correlation functions contribute at higher orders.
The situation seems to be very similar to the Feynman diagram expansion in standard perturbation theory in QFT. In that case we have the relation for the quantum partition function
\be
\log Z[J] = i W[J]\ .
\ee
Here $Z[J]$ is the standard partition function, which is generated by all diagrams. $W[J]$ on the other hand is generated only by connected diagrams. 
Our analysis above suggests that the analogue of connected diagrams in conformal perturbation theory are reduced correlation functions, that is correlation functions for which the contribution of the external field $\varphi$ has been removed from all channels.
To give an idea how this works in a non-rigorous manner, first let us  bring
(\ref{ordern}) to the form
\be\label{ordern2}
I_n :=\frac{\lambda^n}{n!}\int d^2w_1 \ldots d^2 w_n \langle \varphi|\Phi(w_1)\ldots \Phi(w_n)|\varphi\rangle\ .
\ee
We have dropped all prefactors coming from the coordinate transformation. Formally (\ref{ordern2}) looks completely independent of $z_{12}$. However, in the process we roughly rescaled $\varepsilon \to \varepsilon/|z_{12}|$. The advantage of this is that we only need to look for $\log \varepsilon$ divergences, which are then guaranteed to come with the correct $\log |z_{12}|$ contributions.
Next we insert a complete set of states $\phi$ between each $\Phi$,
\be
\frac{\lambda^n}{n!}\sum_{\phi_i}\int d^2w_1 \ldots d^2 w_n \langle \varphi|\Phi(w_1)|\phi_1\rangle\langle\phi_1|\Phi(w_2)|\phi_2\rangle\ldots \langle \phi_{n-1}|\Phi(w_n)|\varphi\rangle\ .
\ee
Note that we can ignore radial ordering of the $\Phi(w_i)$ here, since in the end the factors of course permute.
We then pick out the contribution of the term with $\phi_i=\varphi$. For the term where all $\phi_i=\varphi$, we find
\be
\frac{\lambda^n}{n!} I_1^n = \frac{\lambda^n}{n!} \left( 2\pi C_{\varphi\varphi\Phi} \log \varepsilon^{-2} \right)^n\ .
\ee
Again, this term corresponds to the $n^{\text{th}}$ term in the expansion of $\exp(-2\log(z_1-z_2) h^{(1)})$ in (\ref{shiftexp}). This explains how the higher order terms of the $h^{(1)}$ come about in higher order perturbation theory. It seems likely that a similar pattern also appears in the expansion of the terms of higher $h^{(n)}$.

\bibliographystyle{JHEP}
%%\small\baselineskip=.87\baselineskip
%%\let\bbb\bibitem\def\bibitem{\itemsep1.5pt\bbb}
\bibliography{refmain}

\end{document}